\documentclass[12pt]{article}
\usepackage{amsfonts}
\usepackage{amsmath}
\usepackage{amssymb}
\usepackage[dvips]{color}
\usepackage{epsfig}
\usepackage{graphicx}
\usepackage{slashbox}
\begin{document}

\title{Unparticle-Induced Lepton Flavor Violating Decays \\$\tau\to \ell
(V^0, ~P^0)$}

\author{Zuo-Hong Li\footnote{Email:lizh@ytu.edu.cn},~~ Ying Li\footnote{Email:liying@ytu.edu.cn}~~
and~~Hong-Xia Xu
 \\
{\it  Department of Physics, Yantai University, Yantai 264005,
China}}

\maketitle

\begin{abstract}

We make an evaluation of the lepton flavor violating (LFV) decays
$\tau \to \ell (V^0, P^0)$, where $\ell=e$ or $\mu$ and $V^0(P^0)$
is a neutral vector (pseudo-scalar) meson, in the context of
unparticle physics. The constraints are investigated systematically
on the related coupling parameters from all the available
experimental data, and the parameter values are specified
appropriately. The results show that whereas over the whole
parameter space allowed by experiments all the $\tau \to \ell P^0$
modes have a branching ratio too small to be measurable
experimentally, in a large subspace as observed all of the $\tau \to
\ell V^0$ modes get simultaneously a branching ratio as high as
${\cal O}(10^{-10}-10^{-8})$, which is reachable at the LHC and
super B factory. The important implications are drawn.
\end{abstract}

~~~{\small{\it PACS numbers}:~12.60.-i; 13.00.00; 13.35.Dx; 14.80.-j}\\

~~~{\small{\it Keywords}: Unparticle; Lepton flavor violating
decays; $\tau$ lepton}
\newpage
\section{Introduction}
~~~~~In the Standard Model(SM), massless neutrinos of different
families are not mixed so that lepton flavors are made exactly
conservative, or speaking, a lepton flavor violating (LFV) mode is
forbidden absolutely in the SM. If the neutrino oscillation
phenomenon takes place actually, we can affirm that the neutrinos
are of a nonvanishing mass and thus lepton flavor conservation would
be broken. Even so, LFV processes are still highly suppressed
because of the smallness of neutrino masses. Hence, any distinct LFV
signal can be deemed an indication of new physics beyond the SM.
Recently, there has been an increasing interest in LFV physics. The
current status of this subject is reviewed in \cite{Raidal}.

Given the fact that the operators responsible for LFV transitions
could be provided by most of the existing models beyond the SM, LFV
phenomena could be explored in various theoretical frameworks. Most
of efforts have been devoted to an investigation about LFV decays
\cite{Lfvnp, Mssm, Sseesaw, Gut, IIIseesaw, Aliev, Ding, Hektor} and
lepton anomalous magnetic moments $(g-2)$ \cite{Hektor, Liao1, Yuan,
Luo, Hur, Npam}. A large branching ratio is predicted for $\tau\to
3\ell$, $\ell\gamma$ and $\ell (V^0, P^0)$ (where $\ell=e$ or $\mu$
and $V^0 (P^0)$ is a neutral vector (pseudo-scalar) meson) in some
models such as the MSSM framework \cite{Mssm}, SUSY seesaw mechanism
\cite{Sseesaw}, SUSY-GUT scenario \cite{Gut} and type-III seesaw
model \cite{IIIseesaw}. Of all the existing discussions on LFV,
those based on unparticle theory \cite{Georgi} are especially
intriguing, because LFV processes can proceed at a tree level in
this approach. The phenomenological implications of unparticle
physics have been discussed intensely for the LFV transitions
$\mu\to 3e$ \cite{Aliev}, $\mu \to e\gamma$ \cite{Ding} and $\tau\to
\mu\gamma$ \cite{Hektor}, electron and muon $(g-2)$ \cite{Hektor,
Liao1, Yuan, Luo, Hur} and collider physics \cite{Yuan, Collider}.
Besides, the effects of unparticle have been explored on hadronic
processes \cite{ Luo, Mixing, Mohanta, Kim, He, Bdecay}. More
interestingly, the experimental constraints have been investigated
on some of the unparticle coupling strengths and the important
results have been obtained. For a mini-review on unparticle
phenomenology one can be referred to \cite{Cheung}. The recent
progress in unparticle physics can be found in \cite{Georgi1}. On
the other hand, a continuous experimental search has already
performed for various LFV $\tau$ decays. Very recently, an updated
measurement has been reported on $Br(\tau\to \ell V^0)$ \cite{Belle}
and $Br(\tau \to \ell K_s^0)$ \cite{Barbar}. The estimated
experimental upper limits on the branching ratios are in the range a
few $\times (10^{-8}-10^{-7})$ at $90\%$ confidence level, for
$\tau\to 3\ell$, $\ell\gamma$ and $\ell (V^0, P^0)$ \cite{PDG}.
Though no clear signal has been detected in the current extensive
search for LFV decays, it is expected that the future LHC will probe
$\tau\to 3\mu$ and $\ell (V^0, P^0)$ down to the $10^{-8}$ level,
while a sensitivity of $10^{-10}-10^{-9}$ will be reachable for a
search for $\tau\to 3\ell$, $\ell\gamma$ and $\ell (V^0, P^0)$ at
the super B factor \cite{SuperB}.

Motivated by the recent progress in unparticle phenomenology and
good prospect of the experiments on LFV $\tau$ decays, in this
Letter we intend to make an assessment of $\tau\to \ell (V^0,P^0)$
in the context of unparticle physics to understand the possibility
to discover them in the future experimental searches.

This Letter is organized as follows. In the following section, on a
brief introduction of the basic concepts of unparticle physics, we
address the effective models we use for describing unparticle
interactions with the SM particles and make a simple discussion.
Section 3 is devoted to a derivation of decay rates for $\tau \to
\ell (V^0,P^0)$. A detailed parameter discussion and numerical
evaluation is presented in section 4. The final section is reserved
for summary.

\section{Effective Interactions}
~~~~The scale invariance in the conformal field theory, although not
an exact symmetry of nature, might play an important role in
exploring new physics beyond the SM. It prohibits strictly any
particles with a definite nonzero mass from manifesting themselves
and thus is broken in the SM. But there could be a sector, which is
exactly scale invariant and interacts very weakly with SM particles
at a scale much beyond the SM one. On the basis of a previous study
\cite{Banks}, Georgi \cite{Georgi} suggests that there exist, in a
very high energy theory, SM fields and BZ fields with a nontrivial
infrared fixed point. These two sectors interact with each other by
exchanging particles with a very large mass $M_{\cal U}$. Below this
mass scale, the heavy particles can be integrated out, resulting in
the following local interactions:
\begin{eqnarray}
\frac{1}{M_{ \cal U}^{d_{SM}+d_{BZ}-4}}O_{SM}O_{BZ},
\end{eqnarray}
where $O_{SM}$ is a SM operator with mass dimension $d_{SM}$ and
$O_{BZ}$ an operator with mass dimension $d_{BZ}$ built out of BZ
fields. When the energy scale runs down to a certain scale
$\Lambda_{\cal U}$, at which the scale invariance in the BZ sector
emerges, the renormalizable couplings of the BZ fields bring about
dimensional transmutation. Then below this scale the BZ operators
match onto unparticle ones and correspondingly, the interactions in
(1) match onto an effective interaction of the form
\begin{eqnarray}
\frac{C_{ \cal U}\Lambda_{ \cal U}^{d_{BZ}-d_{ \cal U}}}{M_{ \cal
U}^{d_{SM}+d_{BZ}-4}}O_{SM}O_{ \cal U},
\end{eqnarray}
with $C_{\cal U}$ being a coupling coefficient and $d_{ \cal U}$ the
nonintegral number scale dimension of the unparticle operator $O_{
\cal U}$.

Scale invariant unparticle stuff bears the characters strikingly
other than those of ordinary particles. In particular, scale
invariance can be used to fix the two-point functions of unparticle
operators and further their propagators. The resulting propagators
read,
\begin{eqnarray} \int d^{4}x e^{iP\cdot
x}\langle0\mid T[O_{ \cal U}^{\mu}(x)O_{ \cal U}^{\nu}(0)]\mid
0\rangle=i\frac{A_{d_{ \cal U}}}{2\sin(d_{ \cal
U}\pi)}(-g^{\mu\nu}+\frac{P^{\mu}P^{\nu}}{P^2})
(-P^{2}-i\epsilon)^{d_{ \cal U}-2},
\end{eqnarray}
for a transverse vector unparticle, and
\begin{eqnarray}
\int d^{4}x e^{iP\cdot x}\langle 0\mid T[O_{ \cal U}(x)O_{ \cal
U}(0)]\mid 0\rangle=i\frac{A_{d_{ \cal U}}}{2 \sin(d_{ \cal U}\pi)}
(-P^{2}-i\epsilon)^{d_{ \cal U}-2},
\end{eqnarray}
for a scalar unparticle. The coefficient $A_{d_{ \cal U}}$ is given
by
\begin{eqnarray}
A_{d_{ \cal U}}=\frac{16\pi^{5/2}}{(2\pi)^{2d_{ \cal
U}}}\frac{\Gamma(d_{ \cal U}+\frac{1}{2})} {\Gamma(d_{ \cal
U}-1)\Gamma(2d_{ \cal U})}.
\end{eqnarray}

Since the matching procedure from the BZ operators to unparticle
ones is unknown, unparticles may interact with SM particles in many
possible ways. In the present case, we would like to use the
effective coupling forms suggested by Georgi \cite{Georgi}. Then the
interactions of a vector unparticle with the charged leptons can be
expressed uniformly as
\begin{eqnarray}
{\cal L}_E&=&2\Lambda_{\cal U}^{1-d_{\cal U}}
\overline{E}_{L}\gamma_{\mu}{V}_{E}{E}_{L}O^{\mu}_{\cal U}\nonumber\\
&=&2\Lambda_{\cal U}^{1-d_{\cal
U}}~\bordermatrix{\bar{e}_{L},&\bar{\mu}_{L},&\bar{\tau}_{L}}
\gamma_{\mu}\bordermatrix{&\cr &\lambda_{ee} &\lambda_{e\mu}
&\lambda_{e\tau}\cr &\lambda_{\mu e} &\lambda_{\mu\mu}
&\lambda_{\mu\tau}\cr &\lambda_{\tau e} &\lambda_{\tau\mu}
&\lambda_{\tau\tau}\cr } \bordermatrix{&\cr &e_{L}\cr &\mu_{L}\cr
&\tau_{L}\cr}O^{\mu}_{\cal U},
\end{eqnarray}
where a left-hand lepton vector $E_L$ is introduced, and all the
related coupling constants $\lambda_{ij}$ are arranged in a $3\times
3$ matrix $V_E$ and are treated as a real number. These coupling
constants are in general viewed as a free parameter. A hierarchical
relation, however we may conceive, does exist among some of them,
because the LFV operators might be suppressed to a different degree
by a small factor. We postulate that the following relations are
respected: $\lambda_{\tau\tau}\geq\lambda_{\tau\mu}\geq\lambda_{\tau
e}$ and $\lambda_{\mu\mu}\geq\lambda_{\mu e}$. In fact, such
relations could be accommodated by the existing experimental data,
as will be seen later.

The unparticle interactions with quarks could be discussed in
parallel. Since only three light quarks are involved in the present
situation, it suffices that we confine ourself to the former two
generations. We have
\begin{eqnarray}
&&{\cal L}_U=2\Lambda_{\cal U}^{1-d_{\cal U}}\overline{
U}_{L}\gamma_{\mu}{V}_{U}U_{L}O^{\mu}_{\cal U}\nonumber\\
&&~~~~=2\Lambda_{\cal U}^{1-d_{\cal U}}~\bordermatrix{\bar{
u}_{L},&\bar{c}_{L}} \gamma_{\mu}\bordermatrix{&\cr &\lambda_{uu}
&\lambda_{uc} \cr &\lambda_{cu} &\lambda_{cc} \cr }
\bordermatrix{&\cr &u_{L}\cr &c_{L}\cr}O^{\mu}_{\cal U},
\end{eqnarray}
\begin{eqnarray}
&&{\cal L}_D=2\Lambda_{\cal U}^{1-d_{\cal
U}}\overline{D}_{L}\gamma_{\mu}{V}_{D}{D}_{L}O^{\mu}_{\cal U}\nonumber\\
&&~~~~=2\Lambda_{\cal U}^{1-d_{\cal
U}}~\bordermatrix{\bar{d}_{L},&\bar{s}_{L}}
\gamma_{\mu}\bordermatrix{&\cr &\lambda_{dd} &\lambda_{ds} \cr
&\lambda_{sd} &\lambda_{ss} \cr } \bordermatrix{&\cr &d_{L}\cr
&s_{L}\cr}O^{\mu}_{\cal U},
\end{eqnarray}
which describe the unparticle interactions with up-and down-type
quarks, respectively. For the related flavor conserving couplings
$\lambda_{qq}$ $(q=u,d,s)$ and flavor changing one $\lambda_{sd}$,
we assume them to comply with the numerical relationship
$\lambda_{uu}\sim \lambda_{dd}\sim \lambda_{ss}\geq\lambda_{sd}$.

Correspondingly, the effective interactions involving scalar
unparticle are of the following forms:
\begin{eqnarray}
&&{\cal L}_E'=2\Lambda_{\cal U}^{-d_{\cal
U}}~\bordermatrix{\bar{e}_{L},&\bar{\mu}_{L},&\bar{\tau}_{L}}
\gamma_{\mu}\bordermatrix{&\cr &\lambda'_{ee} &\lambda'_{e\mu}
&\lambda'_{e\tau}\cr &\lambda'_{\mu e} &\lambda'_{\mu \mu}
&\lambda'_{\mu \tau}\cr &\lambda'_{\tau e} &\lambda'_{\tau \mu}
&\lambda'_{\tau \tau}\cr } \bordermatrix{&\cr &e_{L}\cr &\mu_{L}\cr
&\tau_{L}\cr}\partial^{\mu}O_{\cal U},\\
&&{\cal L}_U'=2\Lambda_{\cal U}^{-d_{\cal
U}}~\bordermatrix{\bar{u}_{L},&\bar{c}_{L}}
\gamma_{\mu}\bordermatrix{&\cr &\lambda'_{uu} & \lambda'_{uc} \cr
&\lambda'_{cu} &\lambda'_{cc} \cr }
\bordermatrix{&\cr &u_{L}\cr &c_{L}\cr}\partial^{\mu}O_{\cal U},\\
&&{\cal L}_D'=2\Lambda_{\cal U}^{-d_{\cal
U}}~\bordermatrix{\bar{d}_{L},&\bar{s}_{L}}
\gamma_{\mu}\bordermatrix{&\cr &\lambda'_{dd} &\lambda'_{ds} \cr
&\lambda'_{sd} &\lambda'_{ss} \cr } \bordermatrix{&\cr &d_{L}\cr
&s_{L}\cr}\partial^{\mu}O_{\cal U},
\end{eqnarray}
where it can likewise be assumed that there are the coupling
hierarchies which are of the same structures as the corresponding
ones suggested in the vector unparticle cases, and it should be
understood that the scale dimensions have been set identical for the
two different types of unparticles.
\section{Calculation of Decay Rates  }
~~~~Now we embark upon calculating the decay rates for $\tau\to \ell
(V^0, P^0)$ with the effective interactions $(6)-(11)$. It is easily
noticed that the scalar (vector) unparticle does not couple with a
single vector (pseudo-scalar) meson. Then the decays $\tau\to \ell
V^0$ proceed via just the vector unparticle, while the $\tau\to \ell
P^0$ transitions do by only the scalar unparticle.

\begin{figure}[hbtp]
\begin{center}
\includegraphics[width=0.6\textwidth]{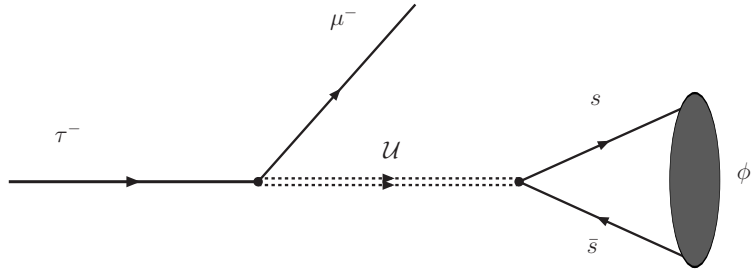}
\caption{\small Feynman diagram of the vector unparticle-induced LFV
decay $\tau\to \mu\phi$.} \label{diagram0}
\end{center}
\end{figure}

In order to discuss the vector unparticle mediated decays $\tau\to
\ell V^0$, we could take $\tau\to \mu\phi$ as an illustrative
example. From the Feynman diagram plotted in Fig.~\ref{diagram0}, we
can write down the transition amplitude as
\begin{eqnarray}
&&{\cal A}(\tau\to \mu \phi)=\frac{\lambda_{\tau\mu}}{\Lambda_{{\cal
U}}^{d_{{\cal U}}-1}}\overline{\mu}
\gamma_{\mu}(1-\gamma^{5})\tau\frac{iA_{d_{ \cal U}}}{2\sin(d_{ \cal
U}\pi)}(-g^{\mu\nu}+\frac{P^{\mu}P^{\nu}}{P^2})
(-P^{2}-i\epsilon)^{d_{ \cal U}-2}\nonumber
\\
&&~~~~~~~~~~~~~~~~~~~\times\frac{\lambda_{ss}}{\Lambda_{{\cal
U}}^{d_{{\cal U}}-1}}
\langle\phi|\bar{s}\gamma_{\nu}(1-\gamma^{5})s|0\rangle,
\end{eqnarray}
where $P^\mu$ is the four momentum of the unparticle, and we have
employed the ideal mixing scheme for the $\omega-\phi$ system.
Further, the above expression can be simplified as,
\begin{eqnarray}
{\cal A}(\tau\to \mu
\phi)=i\frac{\lambda_{\tau\mu}\lambda_{ss}}{2\Lambda^{2(d_{{\cal
U}}-1)}}\frac{A_{d_{\cal U}}}{\sin(d_{\cal U}\pi)}
m_{\phi}f_{\phi}(-m_\phi^{2}-i\epsilon)^{d_{ \cal U}-2}
\overline{\mu} \gamma^{\mu}(1-\gamma^{5})\tau\varepsilon_{\mu}^*,
\end{eqnarray}
using the standard definition $\langle \phi|\bar{s}\gamma_\mu
s|0\rangle= m_{\phi} f_{\phi}\varepsilon_\mu^*$, with $m_{\phi}$,
$f_{\phi}$ and $\varepsilon$ being the corresponding mass, decay
constant and polarization vector, respectively. After summing over
the spins of the final states and averaging over the spins of the
initial state, the decay width is derived as
\begin{eqnarray}
\Gamma(\tau\to \mu \phi)=\frac{|\vec{p}|}{16\pi
m_{\tau}^2}\sum_{spin}|\mathcal{A}(\tau\to \mu \phi)|^{2},
\end{eqnarray}
where $\vec{p}$ stands for the momentum of the outgoing particles in
the $\tau$ rest frame, and
\begin{eqnarray}
\sum_{spin}|\mathcal{A}(\tau\to \mu\phi)|^{2}=\left[\frac{
\lambda_{\tau\mu}\lambda_{ss}}{\Lambda^{2(d_{{\cal U}}-1)}}\frac
{A_{d_{\cal U}}}{\sin(d_{\cal U}\pi)}
m_{\phi}f_{\phi}|(-m_\phi^{2}-i\epsilon)^{d_{ \cal
U}-2}|\right]^{2}\left[
\frac{m_\tau^4}{m_\phi^2}+m_\tau^2-2m_\phi^2\right].
\end{eqnarray}

Using $(13)-(15)$ and making a simple algebraic manipulation, the
decay rates for the other $\tau\to \ell V^0$ modes are easily
achieved. Here we do not give them any more.

For the scalar unparticle mediated decays $\tau \to \ell P^0$, the
hadronic matrix elements $\langle P^0(p)|\bar
q_1\gamma_{\mu}\gamma_5 q_2|0\rangle$ enter into the expressions for
the decay amplitudes. As usual, in the $\pi^0$ and $K^0 (\bar
{K}^0)$ case these matrix elements are parameterized, for instance,
as
\begin{eqnarray}
 \langle K^0 (p)|\bar d\gamma_{\mu}\gamma_5
 s|0\rangle=if_{K}p_{\mu}
\end{eqnarray}
with $p_\mu$ and $f_{K}$, respectively, being the four momentum and
decay constant of the $K^0$ meson. In contrast, the $\eta$ and
$\eta'$ situation is much more complicated because of mixing. The
relevant decay constants are defined by
\begin{eqnarray}
 \langle M(p)|\bar{q} \gamma ^\mu
\gamma_5 q|0 \rangle = \frac{i}{\sqrt{2}}f^q_M p^\mu ,\,\,\,
  \langle M(p)|\bar{s}\gamma^\mu\gamma_5 s|0 \rangle =
if^s_M  p^\mu , \label{hi1}
\end{eqnarray}
where $M=\eta$ or $\eta'$ and $q=u$ or $d$. We would like to
consider the $\eta-\eta'$ mixing effect in the Feldmann-Kroll-Stech
(FKS) scheme \cite{FKS}. In this scheme the physical meson states
$|\eta\rangle$ and $|\eta'\rangle$, in term of the parton Fock
states $|\eta_q \rangle =|u \bar{u} + d \bar{d}\rangle/\sqrt{2}$ and
$|\eta_s\rangle =|s \bar{s}\rangle$, are decomposed as
\begin{eqnarray}
\left( \begin{array}{c} |\eta\rangle\\
|\eta'\rangle \end{array} \right) =
\left( \begin{array}{c c} \cos\phi & -\sin\phi\\
\sin \phi & \cos \phi \end{array}\right) \left( \begin{array}{c}
|\eta_q\rangle\\|\eta_s\rangle \end{array} \right),
\end{eqnarray}
where $\phi$ is the mixing angle. Furthermore, by defining the two
basic decay constants $f_q$ and $f_s$ as
\begin{eqnarray}
 \langle \eta_q(p)|\bar{q} \gamma ^\mu
\gamma_5 q|0 \rangle = \frac{i}{\sqrt{2}}f_qp^\mu,\,\,\,
  \langle \eta_s(p)|\bar{s}\gamma^\mu\gamma_5 s|0 \rangle =
if_sp^\mu, \label{hi2}
\end{eqnarray}
we have the following relations:
\begin{eqnarray}
&f^q_\eta = f_q \cos \phi, &~~~~ f^s_\eta = -f_s \sin \phi,
\nonumber \\
& f^q_{\eta'} = f_q \sin\phi, &~~~~  f^s_{\eta'} = f_s\cos\phi.
\end{eqnarray}
With the aid of the data fitting results $f_q/f_{\pi}=1.07$,
$f_s/f_{\pi}=1.34$ and $\phi=39.3^\circ\pm1.0^\circ$, the desired
values of the decay constants $f^q_{\eta}$, $f^q_{\eta'}$,
$f^s_{\eta}$ and $f^s_{\eta'}$ can be achieved \cite{FKS}.

At present, the decay rates for $\tau\to \ell P^0$ could be
calculated with the known decay constants. As in the vector meson
case, we illustrate our findings of $\tau \to \ell P^0$ with the
resulting expression for the decay width in the $\tau\to \mu K^0$
case,
\begin{eqnarray}
\Gamma(\tau\to \mu K^{0})=\frac{|\vec{p}|}{16\pi
m_{\tau}^{2}}\sum_{spin}|\mathcal{A}(\tau\to \mu K^0)|^{2},
\end{eqnarray}
with
\begin{eqnarray}
&&\mathcal{A}(\tau\to\mu K^{0})=
-\frac{\lambda_{\tau\mu}\lambda_{sd}}{2\Lambda^{2d_{{ \cal U}}}}
\frac{A_{d_{ \cal U}}}{\sin(d_{ \cal U}\pi)}f_{K}m_{K}^{2}(-m_K^{2}-i\epsilon)^{d_{ \cal U}-2}\nonumber\\
&&~~~~~~~~~~~~~~~~~~~~~\times\left[m_{\tau}\overline
\mu(1+\gamma^{5})\tau+m_\mu\overline \mu(1-\gamma^{5})\tau\right],
\end{eqnarray}
and
\begin{eqnarray}
&&\sum_{spin}|\mathcal{A}(\tau\to \mu
K^0)|^{2}=\left[\frac{\lambda_{\tau\mu}\lambda_{sd}}{\Lambda^{2d_{{
\cal U}}}}\frac{A_{d_{ \cal U}}}{\sin(d_{ \cal
U}\pi)}f_{K}m_{K}^{2}|(-m_K^{2}-i\epsilon)^{d_{ \cal
U}-2}|\right]^{2}\nonumber\\
&&~~~~~~~~~~~~~~~~~~~~~~~~~~~~\times\Big[(m_{\tau}^{2}+m_{\mu}^{2})(m_{\tau}^{2}+m_{\mu}^{2}-m_K^{2})-m_{\tau}^{2}m_{\mu}^{2}\Big].
\end{eqnarray}

\section{Parameter Discussions and Numerical Evaluations}
~~~~For a quantitative analysis of these LFV decays, we need to make
a detailed discussion about the various parameters involved in the
calculation.

The mass parameters associated with the present calculation have
been well known, the decay constants of the related light mesons
have been determined better experimentally too. All those are listed
in Tab.\ref{constants}.

The non-integral scale dimension $d_{\mathcal U}$ is calculable in
principles, but difficult to estimate in practice. However, it might
be limited to $1<d_{\mathcal U}<2$, which is to be used here, from
the unitarity \cite {Georgi, Grinstein} and convergence condition.
As concerns the scale parameter $\Lambda_{\cal U}$, we could let it
range from 1~TeV to a few TeV, because ones expect generally that a
certain new physics, if it exists, should appear at such energy
region.

Our main concern, of course, is how the underlying coupling
constants take their values. For having a knowledge about the
couplings $\lambda_{\tau\mu}$ and $\lambda_{\tau e}$, ones have to
make a correlation discussion for the various LFV processes where
these couplings are involved. Unfortunately, the currently available
experimental data are not sufficient to provide them with a decisive
parameter space. For the relevant unparticle-quark couplings, though
we can extract them from the experimental measurements on some
hadronic processes in a certain data fitting way, the uncertainties,
among other things, in the long distant QCD parameters would affect
greatly the accuracy of extraction.

As the case stands, it is needed to work at a level of order of
magnitude, as we make a choice of parameter sets from the regions
allowed experimentally. Before starting our discussion, a couple of
explanations are in order: (1)~Both vector and scalar unparticles
could in general be responsible for a LFV transition. Including
simultaneously contributions of both the unparticles can make not
only the results have a large uncertainty but also the calculation
extremely complicated. In the following parameter discussion we
assume that they two contributes separately, as done in many
studies, and consider only the vector unparticle cases. Also, we
suppose that the corresponding coupling strengths are the same for
the scalar and vector unparticle interactions. (2)~We know that the
unparticle parameters of an effective interaction contains
$\Lambda_{\cal U}$, $d_{\cal U}$ and the coupling constant
$\lambda$. The resulting transition amplitude for a process depends
on the parameter function $f_{d_{\cal U}}(\lambda, \Lambda_{\cal
U})$. With the function values extracted from an experiment, which
are generally relevant to $d_{\cal U}$, the values of $\lambda$ can
be determined at any $d_{\cal U}$, the results being, of course,
dependent on $\Lambda_{\cal U}$. Accordingly, these coupling values,
though changed with $\Lambda_{\cal U}$, correspond to one and the
same $f(\lambda, \Lambda_{\cal U})$ of a fixed value. If we want to
make a theoretical prediction with these extracted coupling
parameters, we could work at an arbitrarily chosen $\Lambda_{\cal
U}$. The final results must have nothing to do with $\Lambda_{\cal
U}$, for the same $f(\lambda, \Lambda_{\cal U})$ enters, which keeps
its value unchanged for different $\Lambda_{\cal U}$. For
convenience, we will work at $\Lambda_{\cal U}=1$ TeV. (3)~The
unparticle couplings $\lambda_{ee}$ and $\lambda_{\mu e}$, as two
important inputs in our parameter determination, have been
investigated in detail in \cite{Liao1, Liao2} and \cite{Ding},
respectively. In the region $1.5\leq d_{\cal U}<2$, the resulting
bounds on $\lambda_{ee}$ and $\lambda_{\mu e}$ are available for the
present case. From the findings obtained by a study on the inviable
positronium decays \cite{Liao2}, we deduce easily that
$\lambda_{ee}\leq 10^{-4}$ for $d_{\cal U}=1.5$, $\lambda_{ee}\leq
10^{-3}$ for $d_{\cal U}=1.6$ and $\lambda_{ee}\leq 10^{-2}$ for
$d_{\cal U}\geq 1.7$. Moreover, $\lambda_{\mu e}$ has a negligibly
small number, as required by the experiments on the $\mu-e$
conversion in heavy nuclei \cite{Ding}, so that we can set it to
zero. These constraint conditions will be used below to restrict
other unparticle couplings. In the region $1<d_{\cal U}<1.5$, the
study indicates that a more stringent restriction on $\lambda_{ee}$
comes from the precise measurement on long-ranged spin-spin
interaction of electrons \cite{Liao2}. However, the results are not
directly applicable and a revaluation is needed. In reality, if we
work in the present context we have to assess not only
$\lambda_{ee}$ but also the related unparticle-quark couplings in
the region $1<d_{\cal U}<1.5$ in which these unparticle parameters
are less known. It is possible to make such an assessment, however
goes beyond the scope of this work. We will choose $1.5\leq d_{\cal
U}<2$ as our work region.

We investigate, to begin with, the possible regions of
$\lambda_{\tau\mu}$ and $\lambda_{\tau e}$ allowed by the existing
experiments \cite{PDG}. The authors of \cite{Hektor} manage to
understand the parameter region of $\lambda_{\tau\mu}$ in a scalar
unparticle model by a combined analysis of the muon $g-2$, $\tau\to
\mu\gamma$ and $\tau\to 3\mu$. They find that as $d_{\cal U} \geq
1.6$ the muon $g-2$ experiment demands that $\lambda_{\mu e}$ have a
negligibly small number, which is in agreement with what is required
by the experiments of the $\mu-e$ conversion in heavy nuclei
\cite{Ding}, and at least one of $\lambda_{\mu\tau,\mu\mu}$ be of
${\cal O}(10^{-1}-1)$. Including further the possible constraints
from the experimental data $Br(\tau\to 3\mu)<3.2\times 10^{-8}$ and
$Br(\tau\to \mu \gamma)<6.8\times 10^{-8}$, they conclude that one
of the two couplings is of ${\cal} O(10^{-1}-1)$, while the other is
at or below order $10^{-2}$. These constraints are possibly weak,
because in the derivation the $\mu$ loop is assumed to dominate in
the $\tau\to \mu\gamma$ transition so that the contribution of the
virtual $\tau$ particle is not included. However, the same
conclusion can yet be drawn in disregard of the constraint of
$\tau\to \mu\gamma$. We make the same investigation within the
present framework by means of the experimental observations of the
muon $g-2$, $\tau\to 3\mu$ and $\tau\to \mu e^+e^-$ (with
$Br(\tau\to \mu e^+e^-)<2.7\times 10^{-8}$), and find that
$\lambda_{\tau\mu}$ can range from ${\cal O}(10^{-3})$ to ${\cal
O}(10^{-2})$ if $\lambda_{\mu\mu}$ takes a larger value of ${\cal
O}(10^{-2}-1)$, and vice versa. From these possible parameter
regions we can pick out our preferred parameter sets:
(1)~$\lambda_{\tau\mu}=10^{-3}$ and $\lambda_{\mu\mu}=10^{-2}$, for
$d_{\cal U}=1.5$. (2)~$\lambda_{\tau\mu}=10^{-3}$ and
$\lambda_{\mu\mu}=10^{-1}$, for $d_{\cal U}=1.6$.
(3)~$\lambda_{\tau\mu}=10^{-2}$ and $\lambda_{\mu\mu}=1$, for
$d_{\cal U}>1.6$. At this point, we must emphasize the fact that the
current experimental data on the tau $g-2$ \cite{PDG} do not provide
more about the couplings involving $\tau$ lepton than we get above
and below, because of the existing sizable uncertainty which allows
us to do theoretical calculation within a considerably large space
of parameter.

As far as $\lambda_{\tau e}$ is concerned, the parameter regions
allowed by $\mu\to e\gamma$ have been evaluated in \cite{Ding},
However, a consistent evaluation requires us to consider a combined
constraint from the processes $\mu\to e\gamma $, $\tau\to 3e$ and
$\tau\to e \mu^+\mu^-$ as well as electron $g-2$. From the
experimental measurements $Br(\mu\to e\gamma)<1.2\times 10^{-11}$,
$Br(\tau\to 3e)<3.6\times 10^{-8}$ and $Br(\tau\to e
\mu^+\mu^-)<3.7\times 10^{-8}$, it follows that $\lambda_{\tau e}$
can be limited to the region $\lambda_{\tau e}\leq {\cal
O}(10^{-4})$ for $d_{\cal U}=1.5$ and $1.6$, while the resulting
upper limits can basically remain at order of $10^{-3}$ for $d_{\cal
U}>1.6$. The constraint is achievable from the electron $g-2$
experiment too, by making a replacement of the corresponding
parameters in the expression for the muon $g-2$ and then confronting
the result with the numerical deviation between the SM estimate and
experimental measurement $|\Delta\alpha| <15\times 10^{-12}$
\cite{electronamg}. But no new results are found. In the numerical
evaluation we will use $\lambda_{\tau e}=10^{-4}$ for $d_{\cal
U}=1.5$ and $1.6$, and $\lambda_{\tau e}= 10^{-3}$ for $d_{\cal
U}>1.6$.

In passing, it is attractive to examine the possible region for
$\lambda_{\tau\tau}$ using the experimental bounds $Br(\tau\to
\mu\gamma)<6.8\times 10^{-8}$ and $Br(\tau\to e\gamma)<1.1\times
10^{-7}$, along with the various constraint conditions obtained
already. The results show that $\tau\to \mu\gamma$ furnishes a
stronger restriction $\lambda_{\tau\tau}\leq {\cal O}(10)$ as
$d_{\cal U}\geq 1.5$ and therefore the possibility of a sizable
unparticle-tau coupling strength cannot be ruled out. Then we can
conclude, according to the present study, that our hierarchy
assumptions $\lambda_{\tau\tau}\geq \lambda_{\tau\mu}\geq
\lambda_{\tau e}$ and $\lambda_{\mu\mu}\geq \lambda_{\mu e}$ are
acceptable at least for the existing LFV experiments. It remains to
be seen whether such relationships are true or not. We can believe
that the future precision measurement on the tau $g-2$ \cite{Tauexp}
would help to clarify this issue.

To turn to the discussion about the unparticle-quark couplings. The
existing constraints on them come mainly from the studies on some
inclusive \cite{He} and exclusive \cite {Mohanta, Kim} decays of B
mesons and neutral meson mixing systems \cite{Luo, Mixing, Mohanta,
Kim}. One expects that the inclusive process $B\to X_s\gamma$ would
provide a stringent constraint on new physics effects, as a result
of the good agreement between the experimental measurement and SM
prediction on $Br(B\to X_s\gamma)$. However, it is not always this
case in the face of unparticle effects \cite{He}. The sensitivity of
$Br(B\to X_s\gamma)$ to the coupling parameters weakens as $d_{\cal
U}>1.5$. As it is, the constraints would become considerably weak as
$d_{\cal U}>1.7$ so that a sizable unparticle-quark coupling
strength is allowed. To have more understanding of the unparticle
parameters, in \cite{Kim} the impacts of unparticle are analyzed on
$B_{d,s}-\bar{B}_{d,s}$ mixing processes and exclusive channels
$B\to \pi\pi,\pi K$, and especially a detailed $\chi^2$ data fitting
is carried out for the $B\to \pi\pi,\pi K$ with the constraints of
$B_{d,s}-\bar{B}_{d,s}$ mixing. The fitting results with $d_{\cal
U}=1.5$ demonstrate that there is a large coupling strength of
${\cal O}(10^n)(n=0, 1)$ for the flavor conserving interactions,
which is compatible with the findings in the $B\to X_s\gamma$
situation. What is particularly interesting is that $\lambda_{uu}$
and $\lambda_{dd}$ turn out to be at the same order of magnitude and
a relatively small number is implied for $\lambda_{sd}$, as expected
by us. It is claimed that with the yielded optimized values of
parameters, the existing various discrepancies may be explained
between the SM predictions and experimental data. We think that
these constraints, though subject to an estimate of uncertainty,
could serve as a valuable reference for us to select proper
parameter values. We assign the following numbers to the related
couplings: $\lambda_{uu}=\lambda_{dd}=\lambda_{ss}=\lambda \sim
10^{-n}$ $(n=-1,0,1,2)$ and $\lambda_{sd}=10^{-n}\lambda$ $(n=1,2)$,
for $d_{\cal U}=1.5$. In the region $d_{\cal U}>1.5$, little is
known about them. Nevertheless, numerous studies show that when
$d_{\cal U}$ increases, the ranges allowed experimentally become
large for unparticle-lepton couplings. The same should be true of
the quark case, for the coupling forms are the same in the two
situations. Taking this point into consideration and for simplicity,
we suggest that these unparticle-quark couplings remain unchanged in
the region $d_{\cal U}\geq 1.5$, due to a certain stringent
restriction condition. This is equivalent to a conservative
estimate. In addition, in the case of $\tau\to \ell(\rho^0,\pi^0)$
we need to confront a combination of two couplings
$\lambda_{uu}-\lambda_{dd}$. Since we are discussing the unparticle
couplings at a level of order of magnitude, it is sound to set
$\lambda_{uu}-\lambda_{dd}$ at the same order as $\lambda_{uu, dd}$.
\begin{table} \caption{\small Summary of the leptonic and hadronic parameters (in units of $\mathrm{MeV}$).}
\begin{center}
\begin{tabular}{cccccccccc}
\hline\hline
 $f_\pi $ & $f_K $ & $f_\rho $ & $ f_\omega $ & $ f_{K^*} $ & $ f_\phi $& $ f_{q} $& $ f_{s} $ \\
 $ 130$&
 $ 160$&
 $ 209$&
 $ 195$&
 $ 217$&
 $ 231$&
 $ 139$&
 $ 174$\\
 \hline
 $m_\pi $ & $ m_K $ & $ m_\rho $ & $ m_\omega $ & $ m_{K^*} $ & $ m_\phi $ & $ m_\eta $ & $ m_{\eta^\prime} $ \\
 $130$&
 $498$&
 $770$&
 $782$&
 $892$&
 $1020$&
 $547$&
 $958$\\
 \hline
  &  &  & $ m_\tau $ & $ m_{\mu} $ & \\
  &  &  &
 $1777$&
 $105$&
 \\
\hline \hline
\end{tabular}\label{constants}
\end{center}
 \end{table}

Now we are a position to make a numerical evaluation. In the first
place, we can notice that for both $\tau \to \ell V^0$ and $\tau \to
\ell P^0$ an approximate order of magnitude relation exists between
the branching ratios, with our selected coupling parameters. In the
$\tau \to \ell V^0$ situation, we have the following observation:
\begin{eqnarray}
&&Br(\tau\to \mu\rho^0)\sim Br(\tau\to \mu\omega)\sim Br(\tau\to \mu\phi) \nonumber\\
~~~~&&~~> Br(\tau\to e\rho^0)\sim
Br(\tau\to e\omega)\sim Br(\tau\to e\phi) \nonumber\\
&&~~~~\geq Br(\tau\to \mu K^{*0}(\bar{K}^{*0}))>Br(\tau\to e
K^{*0}(\bar{K}^{*0})),
\end{eqnarray}
if neglecting the mass difference between muon and electron and
$SU(3)$ breaking effects in the hadron parameters. A similar
relation holds approximately for $\tau\to \ell P^0$. However, it
would suffer from a large $SU(3)$ breaking correction. The numerical
calculations denote that these order of magnitude relations are,
indeed, respected better for $\tau \to \ell V^0$ than for $\tau \to
\ell P^0$.

Let us take a closer look at the behaviors of the branching ratios
in the parameter spaces we adopt. It is clearly seen that the
parameter region $\lambda \leq 1$ is allowed by the experiments,
while the region $\lambda >1$, where the branching ratios for all
the $\tau \to \mu V^0$ go beyond their experimental upper limits, is
prohibited. The allowable parameter sets are fixed as: (I)
$\lambda=10^{-2}$, $1.5\leq d_{\cal U}<2$; (II) $\lambda=10^{-1}$,
$1.55\leq d_{\cal U}<2$; (III) $\lambda=1$, $1.85 < d_{\cal U}<2$.
Over these parameter areas all the $\tau\to \ell P^0$ modes show a
branching ratio less than ${\cal O}(10^{-20})$, which is far from
the experimental reach. We will focus our discussion on the $\tau\to
\ell V^0$ case. For the set I, the branching ratios are of orders
$10^{-14}-10^{-9}$ for $\tau\to \mu V^0$, compared with the
numerical region for $Br(\tau\to e V^0)$ $10^{-16}-10^{-11}$. In the
set II case, whereas the $\tau\to \mu V^0$ modes have a branching
ratio ranging from $10^{-12}$ to $10^{-8}$, the numerical results
for $Br(\tau\to e V^0)$ are located between $10^{-14}$ and
$10^{-9}$. If the set III is used, the numerical values for
$Br(\tau\to \mu V^0)$ vary from $10^{-10}$ to $10^{-8}$, while those
for $Br(\tau\to e V^0)$ do between $10^{-12}$ and $10^{-10}$. To
illustrate the dependence of $Br(\tau\to \ell V^0)$ on $\lambda$ and
$d_{\cal U}$, we can typically consider the $\tau\to \mu\phi$ case
in which the behaviors are shown of $Br(\tau\to \mu\phi)$ in some
parameter regions in Fig.2. Albeit the branching ratios turn out to
be sensitive to $d_{\cal U}$ and $\lambda$, there is still a large
parameter region, as will be seen, in which for any $d_{\cal U}$
almost all the $\tau\to \mu V^0$ modes have a branching ratio as
large as ${\cal O}(10^{-10}-10^{-8})$, which are expected to be
reachable at the LHC and super B factor.

The parameter regions experimentally favorite can be summarized as:
(I) $\lambda=10^{-2}$, $1.5\leq d_{\cal U}<1.8$; (II)
$\lambda=10^{-1}$, $1.55\leq d_{\cal U}\leq 1.9$; (III) $\lambda=1$,
$1.85 <d_{\cal U}<2$. In these regions of the allowable parameter
space, as a matter of fact, all the $\tau\to \mu V^0$ modes except
$\tau\to \mu K^{*0}(\bar{K}^{*0})$ are accessible experimentally,
and in some subregions the same observations can be obtained for
$\tau\to \mu K^{*0}(\bar{K}^{*0})$ and $\tau\to e(\rho^0,
\omega,\phi)$. Only $\tau\to e K^{*0}(\bar{K}^{*0})$ exhibits a
branching ratio below ${\cal O}(10^{-10})$. The partial findings
from these parameter regions, together with the current experimental
upper limits on them, are collected in Tab.2.
\begin{tiny}
\begin{table}[h]
\begin{center}
\setlength{\tabcolsep}{0.08cm} \caption{\small Some selected
numerical results for $Br(\tau\to\ell V^0)$. The corresponding
parameter sets $(\lambda_{\tau\mu},~\lambda_{\tau e}, ~\lambda,
~\lambda_{sd})$ are
$(10^{-3},~10^{-4},~10^{-2},~10^{-3})$,~$(10^{-3},~10^{-4},~10^{-1},~10^{-2})$,
~$(10^{-2},~10^{-3},~10^{-1},~10^{-2})$ and $(10^{-2},~10^{-3},~1,
~10^{-1})$, respectively, for $d_{\cal U}=1.5,~1.6,~1.7$ and 1.8,
and 1.9.} \vspace{0.5cm}
\begin{tabular}{|c|c|c|c|c|c|c|}
\hline \backslashbox{$Mode$}{$d_{\mathcal{U}}$}
   & $1.5$& $1.6$&$1.7$&$1.8$&$1.9$&$$EXP.UL$$\\
   \cline{1-7}
   $\tau\rightarrow \mu\rho^{0}$&$8.0\times10^{-9} $& $2.6\times10^{-8} $&$6.3\times10^{-8} $
    &$4.3\times10^{-9} $&$3.6\times10^{-8} $
    &$6.8\times10^{-8} $\\

   $\tau\rightarrow\mu\omega$&$6.9\times10^{-9} $& $2.2\times10^{-8}
    $&$8.1\times10^{-8} $
    &$3.8\times10^{-9}$&$3.2\times10^{-8} $
    &$8.9\times10^{-8} $\\

   $\tau\rightarrow \mu\phi$&$8.1\times10^{-9} $&$2.9\times10^{-8}
    $&$1.2\times10^{-7} $
    &$6.1\times10^{-9} $&$5.7\times10^{-8} $
    &$1.3\times10^{-7} $\\

   $\tau\rightarrow\mu K^{*0}$&$1.2\times10^{-10} $& $4.2\times10^{-10}
    $&$1.6\times10^{-9} $
    &$7.9\times10^{-11} $&$6.9\times10^{-10} $
    &$5.9\times10^{-8} $\\

   $\tau\rightarrow\mu \bar{K}^{*0}$&$1.2\times10^{-10} $& $4.2\times10^{-10} $&$1.6\times10^{-9} $
    &$7.9\times10^{-11} $&$6.9\times10^{-10} $
    &$1.0\times10^{-7} $\\
   \cline{1-7}
\hline $\tau\rightarrow e\rho^{0}$&$8.1\times10^{-11} $&
    $2.6\times10^{-10} $&$9.4\times10^{-10} $
    &$4.4\times10^{-11} $&$3.6\times10^{-10} $
    &$6.3\times10^{-8} $\\

    $\tau\rightarrow e\omega$&$7.0\times10^{-11} $& $2.2\times10^{-10}
    $&$8.2\times10^{-10} $
    &$3.8\times10^{-11}$&$3.2\times10^{-10} $
    &$1.1\times10^{-7} $\\

    $\tau\rightarrow e\phi$&$8.3\times10^{-11} $&$3.0\times10^{-10}
    $&$1.2\times10^{-9} $
    &$6.2\times10^{-11} $&$5.8\times10^{-10} $
    &$7.3\times10^{-8} $\\

    $\tau\rightarrow e K^{*0}$&$1.2\times10^{-12} $& $4.2\times10^{-12}
    $&$1.6\times10^{-11} $ &$7.9\times10^{-13} $&$7.0\times10^{-12} $
    &$7.8\times10^{-8} $\\

    $\tau\rightarrow e \bar{K}^{*0}$&$1.2\times10^{-12} $&
    $4.2\times10^{-12} $&$1.6\times10^{-11} $
    &$7.9\times10^{-13} $&$7.0\times10^{-12} $
    &$7.7\times10^{-8} $\\
   \cline{1-7}
\end{tabular}
\end{center}
\end{table}
\end{tiny}

So far all the numerical calculations are performed with the fixed
coupling values. However, these parameter values, as has been
emphasized, should are understood as an order of magnitude and thus
we have to consider the effects resulting from the variations of
coupling parameters within their individual orders of magnitude. We
have examined such effects. With the fixed values of
$\lambda_{\tau\mu}$ and $\lambda_{\tau e}$, $\lambda$ dependence of
$Br(\tau\to \mu\phi)$ is plotted in Fig.3.
\begin{figure}[hbtp]
\begin{center}
\includegraphics[width=0.67\textwidth]{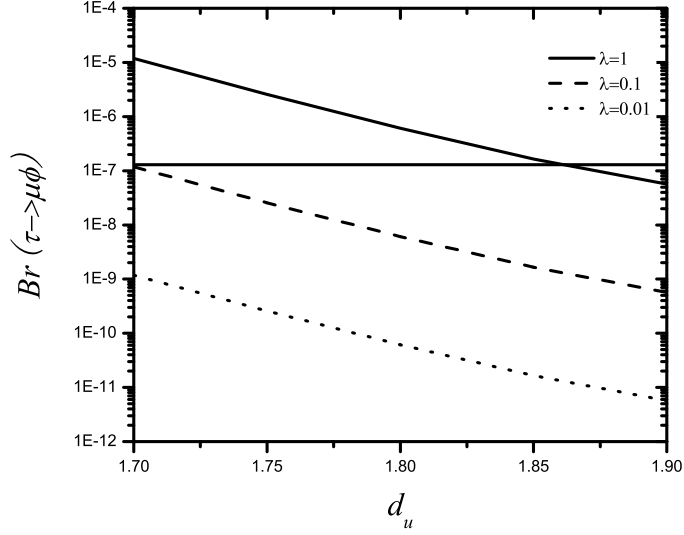}
\caption{\small $d_{\cal U}$ dependence of $Br(\tau\to \mu\phi)$
with the different $\lambda$ values. The horizontal line denotes the
present experimental upper bound on $Br(\tau\to \mu\phi)$.}
\label{diagram1}
\end{center}
\end{figure}
\begin{figure}[hbtp]
\begin{center}
\includegraphics[width=0.67\textwidth]{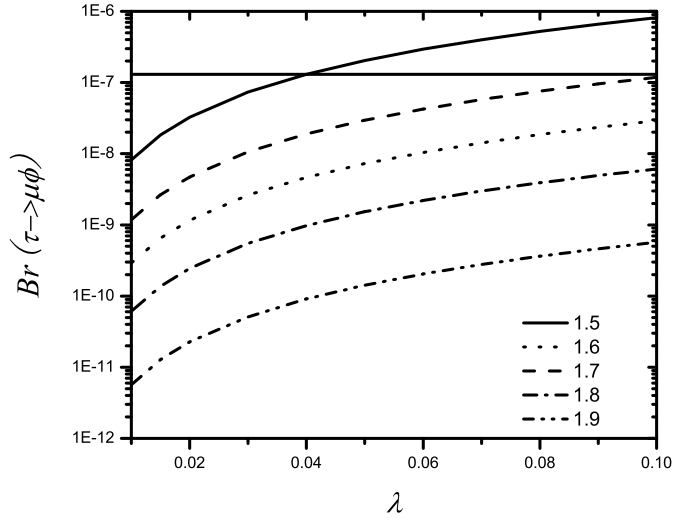}
\caption{\small $\lambda$ dependence of $Br(\tau\to \mu\phi)$ with
the different $d_{\cal U}$. The horizontal line denotes the present
experimental upper bound on $Br(\tau\to \mu\phi)$.} \label{diagram2}
\end{center}
\end{figure}
Obviously, the numerical results can change by up to two orders of
magnitude, when $d_{\cal U}$ remains fixed and $\lambda$ ranges from
$0.01$ to $0.1$. The similar situation appears for $0.1<\lambda \leq
1$. Exploring further the case where all the related parameters vary
simultaneously, we would have a much larger numerical range. But
this also indicates that there could are more theoretical results
which are within the experimental reach. Those listed in Tab.2 are
only some estimated lower bounds on $Br(\tau\to \ell V^0)$. In fact,
when the couplings $\lambda_{\tau\mu}$ and $\lambda_{\tau e}$ change
at the same time within their respective ranges, we can work within
the expanded parameter regions for $\lambda$ and $d_{\cal U}$: (I)
$0.01\leq \lambda < 0.04$, $d_{\cal U}=1.5$; (II) $0.01\leq \lambda
< 0.1$, $1.5<d_{\cal U}<2$; (III) $\lambda=0.1$, $1.55\leq d_{\cal
U}<2$; (IV) $0.1<\lambda<1$, $1.75 < d_{\cal U}<2$; (V) $\lambda=1$,
$1.85 < d_{\cal U}<2$. In these parameter spaces there are many
favorable subspaces as observed, in which all the $\tau\to \ell V^0$
modes can get simultaneously a branching ratio of ${\cal
O}(10^{-10}-10^{-8})$. By contrast, all the $\tau\to \ell P^0$ modes
remain still inaccessible to experiments over these areas.

Conversely, the restrictions can be inspected on the coupling
parameters from the experiment data on $Br(\tau\to \ell (V^0,
P^0))$. By limiting ourself to the parameter area $0.01\leq \lambda
\leq 1$, the constraints on $\lambda_{\tau\mu}$ and $\lambda_{\tau
e}$ are observed to be weaker in a large region of the parameter
spaces for $\lambda$ and $d_{\cal U}$, and rather loose in some
subspaces like $\lambda=0.01$ and $1.6\leq d_{\cal U}<2$, when
compared with those presented above.

Again we stress that all the presented numerical results, despite
achieved at $\Lambda_{\cal U}=1$ TeV, maintain unchanged as
$\Lambda_{\cal U}$ increases, and besides, the scalar unparticle
mediated $\tau\to \ell P^0$ are evaluated using the same scale
dimensions and coupling strengths as in the vector unparticle case.
Even if we regard these corresponding parameters as independent of
each other, it is yet difficult to get a interesting result for
$\tau\to \ell P^0$, since it is hardly conceivable that the related
coupling constants have a considerably sizable number in such a
case. Once the coupling parameters become better understood in the
whole $ d_{\cal U}$ region $1< d_{\cal U} <2$, we could make a more
complete and reliable assessment of these LFV processes. But it
seems likely that with the parameter sets specified adequately the
hierarchical relation $Br(\tau\to\ell V^0) \gg Br(\tau\to \ell P^0)$
will be kept valid, although the branching ratios alter with change
in parameter values. Of course, to do calculation with different
unparticle coupling scenarios would in general lead to different
results. It is desirable to enquire into these LFV processes in
other unparticle coupling schemes.

Our findings for $\tau\to \ell V^0$ appear to be comparable with
some of the existing estimates \cite{Mssm, Sseesaw, Gut, IIIseesaw}.
Nevertheless, in the $\tau \to \ell P^0$ case we have a branching
ratio much less than those for $\tau\to \ell V^0$, presenting a
striking contrast to the predications of the other approaches.
\section{Summary}
~~~We have made a detailed analysis for the unparticle induced LFV
decays $\tau\to \ell (V^0, P^0)$ in an effective model with a
hierarchical relation suggested among some of the coupling
constants.

To get a consistent and believable assessment, all the available
experimental data have been used to constrain the unparticle
couplings. From the obtained constraint conditions, the parameter
values for the related couplings have been specified appropriately.
As a by-product, it is found that a sizable $\lambda_{\tau\tau}$ is
allowed by the current experimental data, and our hierarchy
hypotheses $\lambda_{\tau\tau}\geq \lambda_{\tau\mu}\geq
\lambda_{\tau e}$ and $\lambda_{\mu\mu}\geq \lambda_{\mu e}$ can be
accommodated by these constraint conditions.

We have evaluated the branching ratios and examined the possibility
to experimentally discover these modes in the near future. In the
parameter region $\lambda > 1$, all the $\tau\to \mu V^0$ modes have
a branching ratio exceeding their individual experimental upper
limits. The experimentally allowed regions for $\lambda$ and
$d_{\cal U}$ are determined approximately as: (I) $0.01\leq\lambda <
0.04$, $d_{\cal U}=1.5$; (II) $0.01\leq\lambda<0.1$, $1.5<d_{\cal
U}<2$; (III) $\lambda=0.1$, $1.55\leq d_{\cal U}<2$; (IV)
$0.1<\lambda<1$, $1.75 < d_{\cal U}<2$; (V) $\lambda=1$, $1.85 <
d_{\cal U}<2$. In many regions of these parameter spaces, for all
the $\tau\to \ell V^0$ modes we can have simultaneously a branching
ratio of orders $10^{-10}-10^{-8}$, which are expected to be
accessible at the LHC and super B factory. Compared with the
$\tau\to \ell V^0$ case, all the $\tau\to \ell P^0$ modes show a
branching ratio beyond the experimental reach.

Also, we have inspected the limits imposed on the couplings
$\lambda_{\tau\mu}$ and $\lambda_{\tau e}$ by the experiments on
$\tau\to \ell (V^0, P^0)$, observing that there is a looser bound
than those yielded by the other available LFV experiments, in a
large subspace of the parameter spaces $0.01\leq\lambda\leq1$ and
$1.5\leq d_{\cal U}<2$.

It is explicitly too early to draw a final conclusion whether these
LFV decays are observable experimentally. We have to await the
improvement in experiment and progress in unparticle phenomenology.
Different from the predictions of the other new physics models,
however, the unparticle approach gives the numerical relation
$Br(\tau\to \ell V^0) \gg Br(\tau\to \ell P^0)$, with the
implication that there is a greater discovery potential of $\tau\to
\ell V^0$ than that of $\tau\to \ell P^0$ in future experiments. If
this gets confirmed in the future experimental searches, the present
research is perhaps instructive in identifying whether or not these
LFV processes are induced or dominated by unparticles.

\section*{Acknowledgements}

~~~This research is in part supported by the National Science
Foundation of China under Grant Nos. 10675098, 10747156 and
10805037. Y. Li would like to thank Dr. W. Wang for some valuable
discussions.

\end{document}